\documentclass[%
 reprint,
 amsmath,amssymb,
 aps,
]{revtex4-2}
\usepackage{xcolor}

\usepackage{graphicx}% Include figure files
\usepackage{dcolumn}% Align table columns on decimal point
\usepackage{bm}% bold math
\usepackage{appendix} 
\usepackage{xcolor}

\usepackage{tikz}
\usetikzlibrary{arrows.meta}

\definecolor{plum}{rgb}{0.36078, 0.20784, 0.4}
\definecolor{chameleon}{rgb}{0.30588, 0.60392, 0.023529}
\definecolor{cornflower}{rgb}{0.12549, 0.29020, 0.52941}
\definecolor{scarlet}{rgb}{0.8, 0, 0}
\definecolor{brick}{rgb}{0.64314, 0, 0}
\definecolor{sunrise}{rgb}{0.80784, 0.36078, 0}
\definecolor{lightblue}{rgb}{0.15,0.35,0.75}
\definecolor{carolina}{RGB}{153, 186, 221}
\tikzstyle{axisarrow} = [-{Latex[inset=0pt,length=5pt]}]

\begin{document}

\preprint{APS/123-QED}

\title{Slide FFT on a homogeneous mesh in wafer-scale computing}
%\title{High-throughput signal analysis by wafer-scale FFT}

\author{Maurice H.P.M. van Putten}
 \altaffiliation[]{Physics and Astronomy, Sejong University, Seoul, South Korea, and INAF-OAS Bologna, via P. Gobetti, 101, I-40129 Bologna, Italy}
\email{mvp@sejong.ac.kr}

\author{Leighton Wilson}
 \altaffiliation[]{Cerebras, 1237 E. Arques Ave
Sunnyvale, CA 94085}

\author{Adam W. Lavely}
 \altaffiliation[]{Lawrence Berkeley National Laboratory, 1 Cyclotron Rd
 Berkeley, CA 94720}

\author{Mark Hair}
 \altaffiliation[]{Cerebras, 1237 E. Arques Ave
Sunnyvale, CA 94085}


\date{\today}

\begin{abstract}
Searches for signals at low signal-to-noise ratios frequently involve the Fast Fourier Transform (FFT). For high-throughput searches, we here consider FFT on the homogeneous mesh of Processing Elements (PEs) of a wafer-scale engine (WSE). To minimize memory overhead  in the inherently non-local FFT algorithm, we introduce a new synchronous slide operation ({\em Slide}) exploiting the fast interconnect between adjacent PEs. Feasibility of compute-limited performance is demonstrated in linear scaling of Slide execution times with varying array size in preliminary benchmarks on the CS-2 WSE. The proposed implementation appears opportune to accelerate and open the full discovery potential of FFT-based signal processing in multi-messenger astronomy.
\end{abstract}


\maketitle
%\footnote{Draft 1.0 - not for distribution}

\section{Introduction}

The Fast Fourier Transform (FFT) is central to signal processing and high-performance computing \citep{gar09}. 
High-throughput signal analysis hereby critically depends on FFT performance. 
The FFT algorithm optimizes efficiency in compute, notably in the Cooley-Tukey algorithm \citep{coo65} with $5 n \log_2 n$ floating point operations 
(FLOPs) per transform of length $n$ \citep{van02}. This performance bears out well in CPU-based implementations \citep[e.g.][]{fri05,pek12,lee13,pli18,dai03,lu21}. However, real-time performance sensitively depends on memory overhead
\cite[][]{jeo18} which varies with the architecture of the computing device.

Optimizing real-time signal processing has been approached on dedicated hardware including field-programmable gate arrays (FPGAs) \citep[e.g.][]{woo17} and graphics processor units (GPUs) \citep[e.g.][]{brag15}. %{clm15}. 
These approaches can be advantageous in low latency searches for signals at low signal-to-noise ratios (SNR) by FFT-based matched filtering \citep{tur60}.
Filtering a continuous stream in real-time requires a filter throughput faster than the input data rate.
%\begin{eqnarray}
%    {W=N \dot{T}
%    }
%\label{EQN_W}
%\end{eqnarray}
This is particularly opportune in searches of un-modeled transient signals at detector-limited sensitivity from astrophysical gravitational-wave sources in the Local Universe \citep{van19,van23a}. 

\begin{figure}
    \centering
    \includegraphics[scale=0.19]{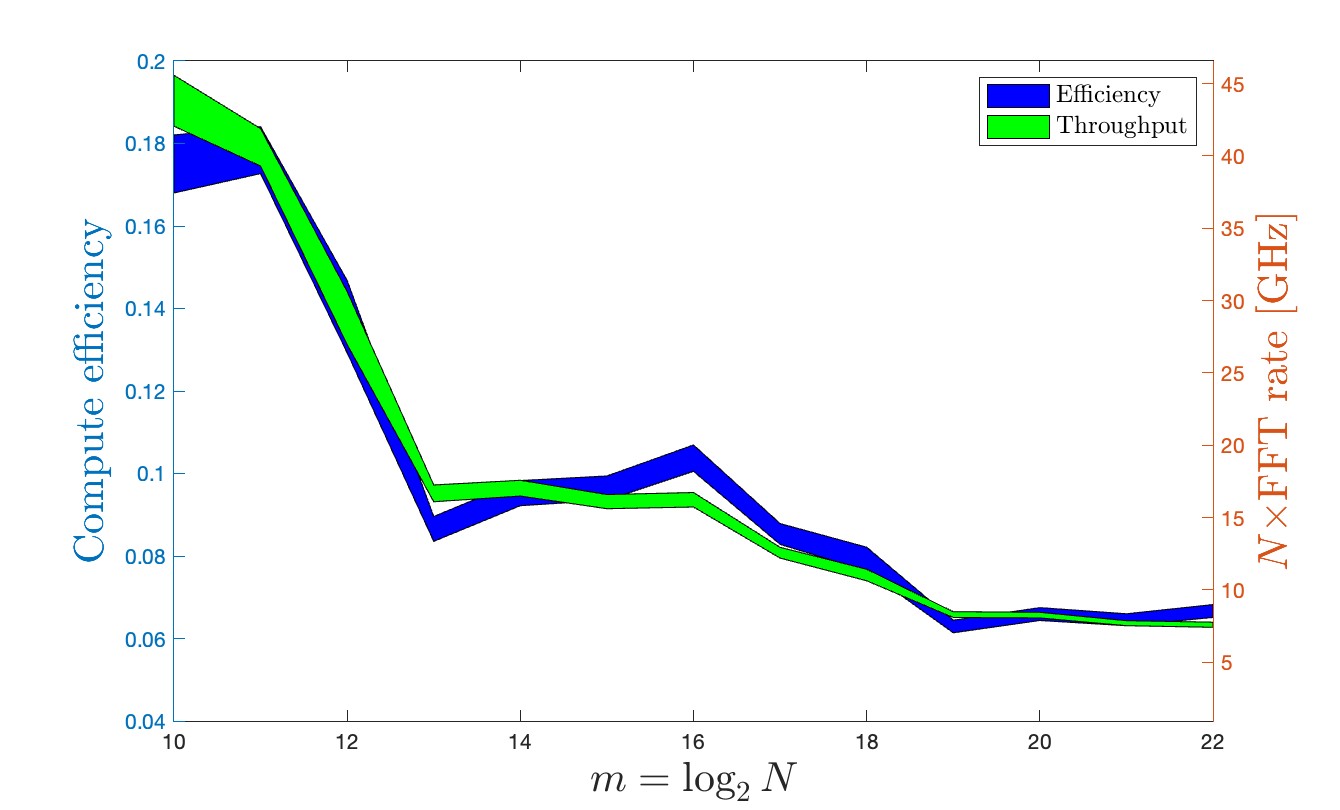}
    \caption{FFT performance on a GPU (Radeon VII with HBM2) by efficiency (left scale) and throughput (right scale) in matched filtering, expressed by the product of transform size $N=2^m$ and FFT transform rate. Results are computed with {\em clFFT} \citep{brag15} in complex single-precision and out-of-place as a function of transform size $N$ in batch mode, shown across memory allocations $M=64$ MB and $M=1024$MB in Global Memory. 
    Efficiency is limited by memory bandwidth especially when transform sizes exceed the size of Local Memory, noticeably across $N=2^{12}$, leaving about 8\% of theoretical peak compute performance in f32. 
    (After Fig. D.1 of \citep{van23a}.)}
    \label{fig0}
\end{figure}

The theoretical limit of FFT compute performance is inevitably below theoretical peak compute-performance due to memory overhead on any architecture. This is not surprising since the Fourier transform is non-local. 
While GPUs offer substantial acceleration over CPUs, performance drops significantly when FFT-array sizes exceed the limits of Local Memory. In fact, the resulting bandwidth-limited performance is significantly below theoretical compute-performance (Fig. \ref{fig0}). 

{\em Wafer-Scale Engines} (WSE) offer a radically new architecture for massively parallel computing on a 
homogeneous mess of Processing Elements (PEs) with fast on-chip interconnect \citep{roc20}.
This appears promising provided memory overhead is kept low relative to compute \citep{li10}.
The Cerebras WSE CS-2 presents a novel architecture in a single chip consisting of 850,000 PEs (Fig. \ref{fig:cs2arch}). 
CS-2 utilizes the maximum silicon square that can be cut from a 300\,mm diameter wafer: a square array of 84 dies over a surface area of 550\,mm$^2$.  The die is interconnected via a custom interconnect by on-chip routers, one per PE. It presents
a homogeneous 2D grid fabric of PEs, each connected to its nearest neighbors North, South, East, and West.
At 850\,MHz, the 850,000 PEs of CS-2 offer a theoretical performance of 240 TFLOP/s at about 3 clock cycles per FLOP.
In FFT transform sizes limited to Local Memory, 3D FFT of size $512^3$ ($\sim 3$\,TFLOPs) is realized in $\sim$\,1\,ms \cite{ore22}.

Here, we study memory overhead in FFT (\S2) on a homogeneous mesh (\S3) by
{\em synchronous slide} of data to facilitate transforms of arbitrary size (\S4).  
The potential of compute-limited performance is demonstrated on the CS-2 by linear scaling of memory 
overhead with data-size (\S5). We summarize our outlook in (\S6).

\begin{figure*}
    \centering
    \includegraphics[scale=0.16]{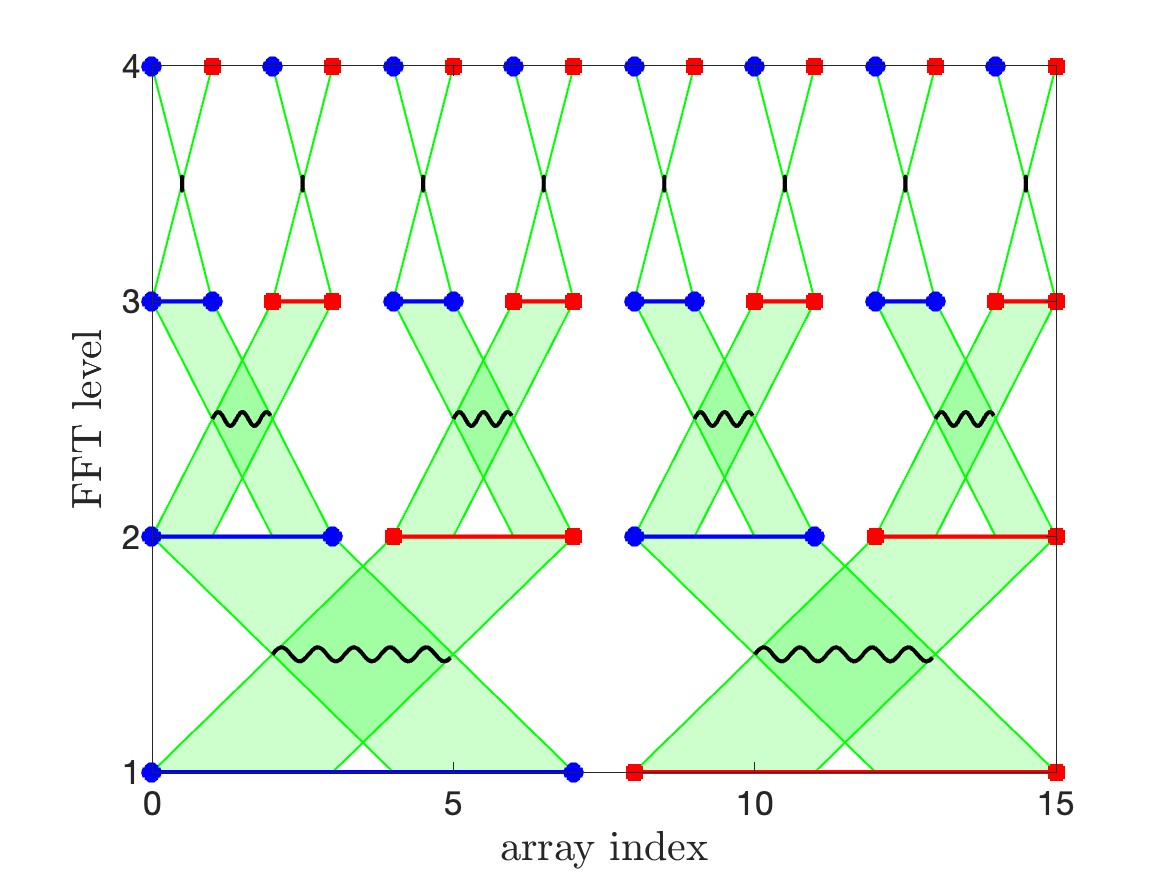} %{plotA-8.1}
    \includegraphics[scale=0.16]{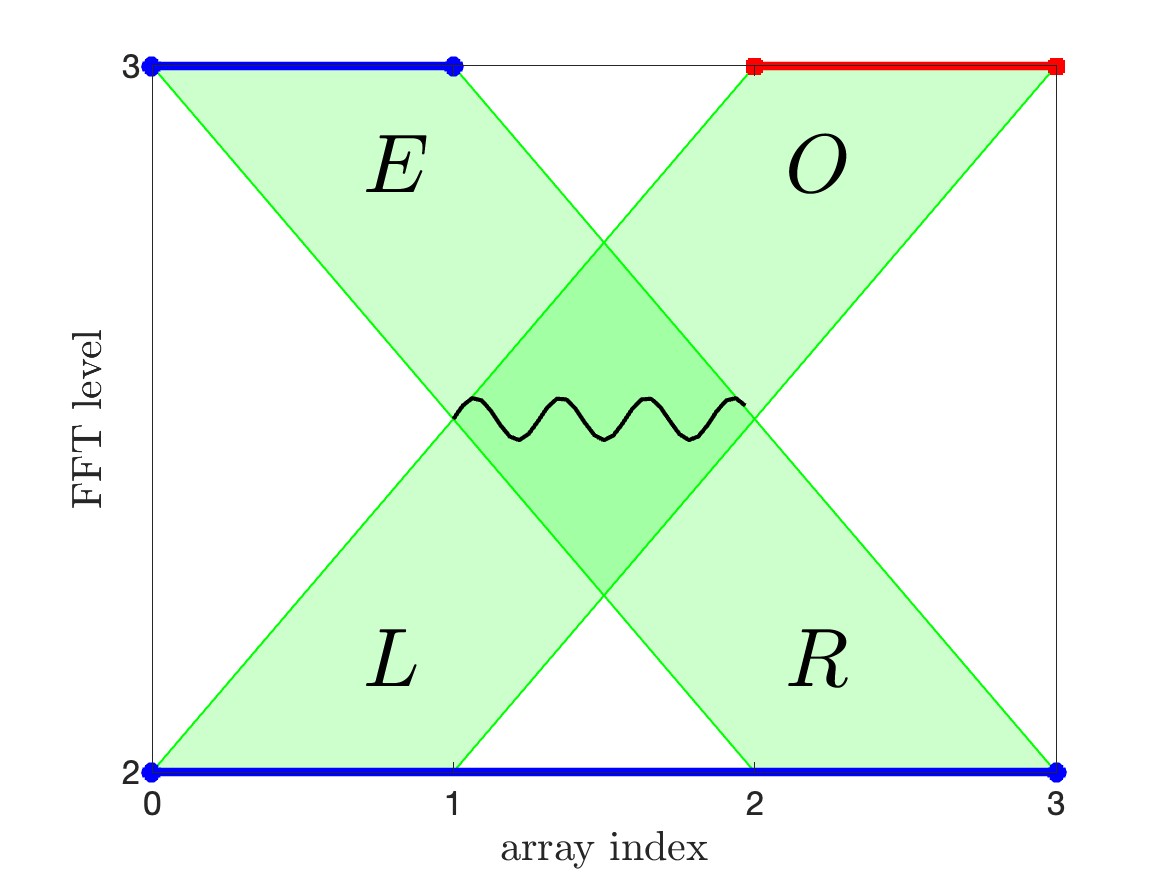}
    \caption{(Left panel.) Following index permutation of input
    data, FFT takes a path in reverse over $m=\log_2n$ levels $p=m,m-1,\cdots,1$. At each $p$, adjacent segments $E$, $O$ are concatenated following a rotation (\ref{EQN_LR}), amenable to embarrassingly parallel computing of $n/2^p$ crossings. 
    (Right panel.) Crossing diagram of rotations (\ref{EQN_EO}) of adjacent segments showing (\ref{EQN_LR}) (wiggly line), 
    subsequent to alignment of $E$ and $O$ in shared memory, followed by sliding output $L,R$ back as input to level $p-1$.
    }
    \label{fig1}
\end{figure*}

\section{Memory overhead in FFT}
%Cooley-Tukey algorithm}

To revisit memory overhead in FFT, we consider an array of samples 
$D=\left[d_0,d_1,\cdots, d_{n-1}\right]$ of
size $n=2^m$. It starts with a permutation of indices produced by sequential partitioning in segments of size $n/2^{p-1}$ over levels $p=1,2,\cdots, m$. Following this pre-processing, FFT is completed over a reverse path by concatenating adjacent elements following crossings on levels $p=m,m-1,\cdots, 1$ (further below).

To illustrate, index permutation is evaluated over $m=\log_2n$ steps (Appendix A), e.g.,
\begin{eqnarray}
 I=\begin{bmatrix}
        0 &      1 &      2 &      3 &      4 &      5 &      6 &      7 \cr
        0 &      2 &      4 &      6 &      1 &      3 &      5 &      7 \cr
        0 &      4 &      2 &      6 &      1 &      5 &      3 &      7
\end{bmatrix}
\label{EQN_1a}
\end{eqnarray} 
for an array of $n=8$ elements over $m=3$ levels. These permutations are readily tabulated, and therefore are not included in evaluation of efficiency. 

FFT proceeds by merging adjacent segments $Y=\left(E~O\right)^T$ of 
size $n/2^{p-1}$ in step $p=m,m-1,\cdots,1$ of even ($E$) and odd ($O$) indexed elements of the array (with indices permuted). 
Merging is a concatenation of the transforms $Y^\prime=\left(L~R\right)$ by a sum and difference,
\begin{eqnarray}
%\begin{array}{l}
L = E + O^\prime, ~~ R = E - O^\prime,
%\end{array}
\label{EQN_LR}
\end{eqnarray}
equivalent to a rotation,
\begin{eqnarray}
Y^\prime = \sqrt{2}\left(\begin{array}{rr}
0 & 1 \\
1 & 0 
\end{array}\right)
{\cal R}_{\pi/4}
\left(\begin{array}{cc}
1 & 0 \\
0 & U_p 
\end{array}\right)Y,
\label{EQN_Y}
\end{eqnarray}
where 
\begin{eqnarray}
{\cal R}_{\pi/4}=\frac{1}{\sqrt{2}}\left(\begin{array}{cr}1 & 1 \\ 1 & -1\end{array}\right).
\end{eqnarray}
Here, the trigonometric part in FFT at level $p$ is in the diagonal matrix 
\begin{eqnarray} 
U_p={\rm dia}\left(1,e^{-i\frac{2\pi}{N}},\cdots,e^{-\frac{2\pi}{N}\left(N/2-1\right)}\right)
\end{eqnarray}
of size $N=2^{m-p+1}$, representing $N/2$ samples on the semicircle of $S^1$ in the lower half-plane with determinant $\left|U_p\right|=1$ $(p=1)$ and $\left|U_p\right|=i$ $(1<p\le m)$.
%\begin{eqnarray}
%U_p={\rm dia}\left(1,e^{-i\frac{2\pi}{N}},\cdots,e^{-\frac{2\pi}{N}\left(N/2-1\right)}\right) 
%\end{eqnarray}
%where $\left(N=2^{m-p+1}\right)$.  $U_p$ represents $N/2$ samples on the semicircle of $S^1$ in the lower half-plane with determinant $\left|U_p\right|=1$ $(p=1)$ and $\left|U_p\right|=i$ $(1<p\le m)$.

\begin{figure}
    \includegraphics[scale=0.16]{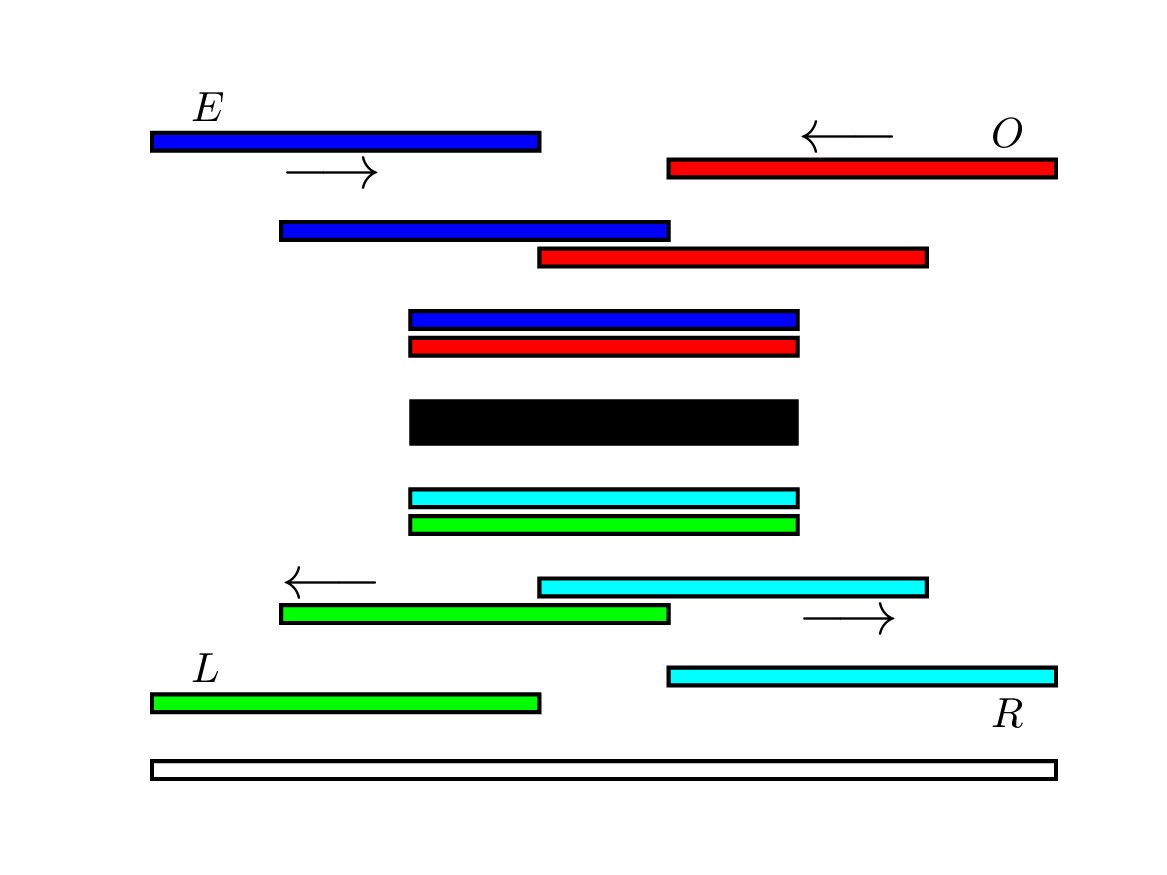}
\caption{
    Memory overhead in aligning $E$ and $O$ in (\ref{EQN_LR}) at level $p$. From top to bottom: 
    coherent shifts to a shared memory location, $E\times O\rightarrow LR$, followed by coherent shifts back to the original location of $E,O$. 
    The concatenated segment (white) is $E$ or $O$ at level $p-1$.   
    }
    \label{fig2}
\end{figure}

Figure \ref{fig1} schematically shows
\begin{eqnarray}
E\times O\rightarrow RL,
\label{EQN_EO}
\end{eqnarray}
aligning $E$ and $O$ to shared memory for compute (\ref{EQN_Y}) and concatenation of output $R, L$. 
The transform (\ref{EQN_EO}) appears one or multiple times in each level $p$ in mutually independent operations, amenable to embarrassingly parallel computing at each transform level (Fig. \ref{fig1}, Appendix A). This is ideally suited for a uniform mesh of PEs {\em provided memory overhead in alignment is negligible} (Fig. \ref{fig2}). 

\section{Performance in Slide FFT}

When data are distributed over a linear array of PEs, Eq. (\ref{EQN_EO}) is non-local and incurs data-transfers between the Local Memory of neighboring PEs. 
In this event, (\ref{EQN_LR}) triggers memory overhead in aligning $E$ and $O$ and in concatenation by {\em sliding}:
\begin{itemize}
\item {\em Forward}: aligning $E$ and $O$ for compute in shared memory location; 
\item {\em Backward}: updating the original memory locations of $E$ and $O$ with output $L$ and $R$, respectively. 
\end{itemize}
It appears natural to slide $E$ and $O$ to a shared memory location half-way in between (shown in Fig. \ref{fig2}), supporting sufficient allocation for $E,O,L$ and $R$ following compute in the rotation Eq. (\ref{EQN_LR}).

Sliding data over a mesh incurs memory overhead that scales with $n/2^{p-1}$ in sliding data. 
Sliding time depends on the velocity of propagation determined by bandwidth between adjacent processors. 
%On a WSE, this incurs a minor latency in {\em ramp up} and {\em down} between local memory and routers of a PE. 
On the CS-2, this incurs a minor latency in sending data up or down a PE's {\em ramp} connecting its CE to its router which communicates with neighboring PEs.
Cumulatively over each transform level $p=m, m-1, \cdots, 1$, communication overhead is $n/2+n/4+\cdots 1 \simeq n$, and double this amount including the sliding of $L,R$ and back to the original locations of $E$ and, respectively, $O$.

Given raw compute performance for multiply/add, compute efficiency subject to memory overhead in data transfer across neighboring PEs depends on the number of clock cycles $a$ per transfer and the associated compute,
\begin{eqnarray}
\alpha = \frac{\#{\rm clock\,cycles\,per\,transfer}}{\#{\rm clock\,cycles\,per\,FLOP}} = \frac{a}{b},
\label{EQN_alpha}
\end{eqnarray}
where $a$ and $b$ are per datum. 
Given aforementioned $5n\log_2n = 5nm$ FLOPs per FFT, compute efficiency of Slide FFT satisfies 
\begin{eqnarray}
{\cal \eta} = \frac{5bmn}{an+5bmn} \simeq 1-\frac{\alpha}{5\,m} + {\cal O}\left(\left(\frac{\alpha}{5\,m}\right)^2\right).
\label{EQN_eta}
\end{eqnarray}

In Slide FFT extending over $2^k$ PEs - a wave of length $k$ on the WSE - the time to align $E$ and $O$ segments (Fig. \ref{fig2}) scales linearly with the number of array elements stored at each processor, e.g., $2^{m-k}$.
For processors with one thread compute time handling Eq. (\ref{EQN_LR}) scales with the same. While total processing time scales inversely with $2^k$, efficiency Eq. (\ref{EQN_alpha}) is invariant under the choice of $k$. 
By (\ref{EQN_eta}), it follows that efficiency of Slide FFT satisfies 
\begin{eqnarray}
\frac{\alpha}{5\,m} \ll 1.
\label{EQN_m}
\end{eqnarray}

\section{The Slide operation}

On a WSE, Eq. (\ref{EQN_LR}) is realized at Eq. (\ref{EQN_m}) by standard operations of floating point multiply/add on data local to each PE and one radically new operation when work space exceeds Local Memory.
Facilitating array sizes in excess of Local Memory, we propose sliding data arrays over a homogeneous mesh of PEs.

Let ${\bf p}$ index the PE positions on a rectangular grid of a WSE. 
We define a Slide of data $A$ at ${\bf p}_0={\&}A$ to $B$ at ${\bf p}_1={\&}B$,
\begin{eqnarray}
B={{*}}{\rm {\bf Slide}}_{{\bf p}}({\&}A),
\label{EQN_s}
\end{eqnarray}
by translation over 
\begin{eqnarray}
{\bf p}={\bf p}_1-{\bf p}_0.
\label{EQN_p}
\end{eqnarray}

The memory overhead Eq. (\ref{EQN_s}) is expected to satisfy linear scaling with size of $A$, distributed over Local Memory of the PEs in a choice of wave length covering the desired FFT transform size. 
This overhead represents a waiting time prior and subsequent to compute in Eq. (\ref{EQN_EO}), 
Ideally, this is overhead is invariant of the choice of wave length $k$, provided it contains sufficient storage in Local Memory aggregated over $k$ PEs. If so, this leaves Eq. (\ref{EQN_alpha}) invariant with respect to $k$.

\section{Slides on the CS-2}

\begin{figure}
    \includegraphics[scale=0.45]{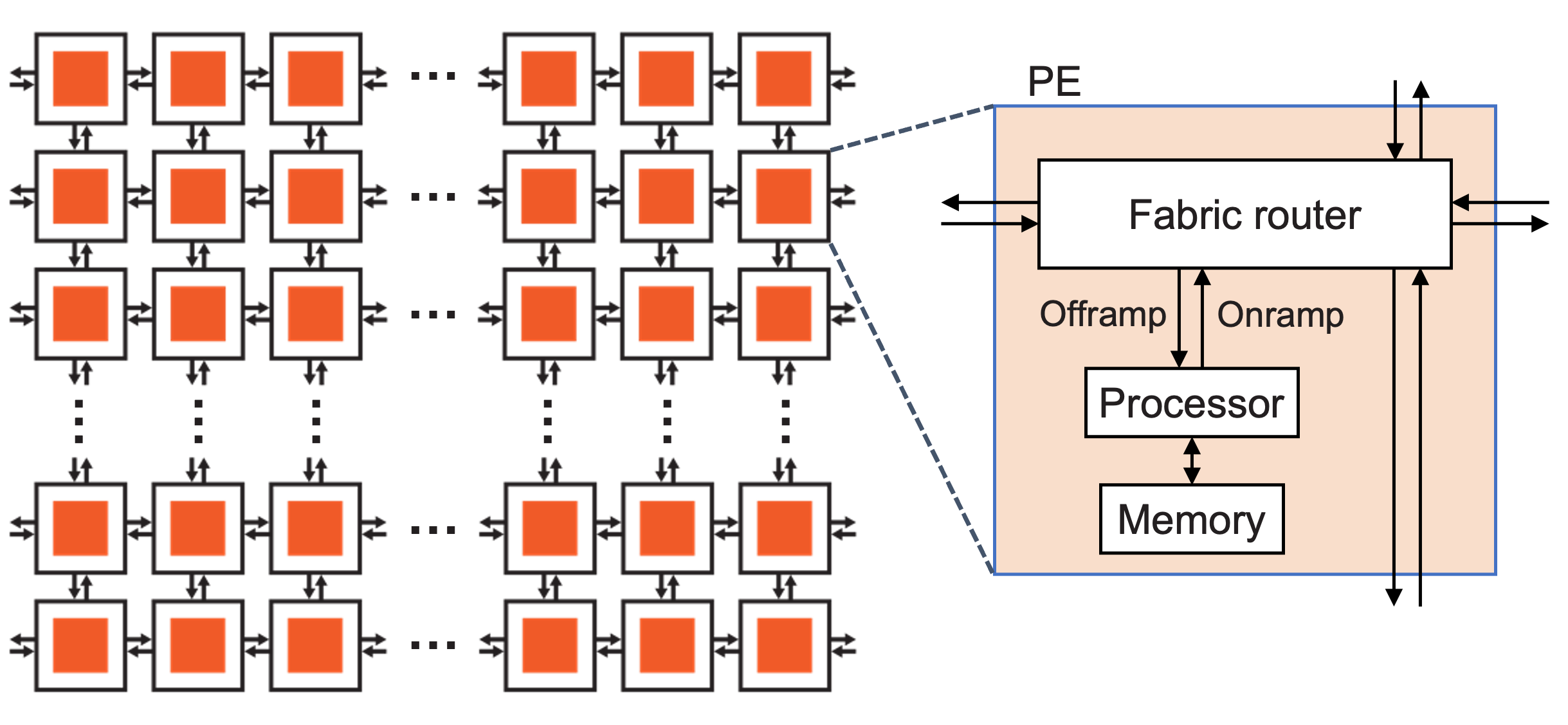}
\caption{Architecture of the wafer-scale engine CS-2 comprising a homogeneous grid of 850,000 PEs (left). Each PE consists of a Compute Element (CE) supported by 48 kB of SRAM Local Memory. Local Memory is connected over a bus to a router enabling bi-directional communication to its nearest neighbors (right). 
    }
    \label{fig:cs2arch}
\end{figure}

On the CS-2, PE performance {\em per clock cycle} is as follows:
\begin{itemize}
    \item {\em Compute:} fused multiply-accomulate (FMAC) in eight 16-bit or two 32-bit FMAC operations by the CE;
    \item {\em Local Memory:} a 128-bit read and a 64-bit write to 48\,kB Local Memory (SRAM);
    \item {\em Router bandwidth:} 32-bit packets (bi-directional) between its nearest neighbors.
\end{itemize}
Additionally, there is a modest latency in communicating between PEs in data-transfer between a CE and its router, i.e., about 2-7 cycles per a 32-bit packet transfer.
Taken together, the above represents a few clock cycles for a floating point operation (FLOP).

\begin{figure*}
    %\includegraphics[scale=0.5]
    %{cycle_per_elem_sim.png} \\
    %(a) \\
    %\includegraphics[scale=0.5]
    %{cycle_per_elem_cs2.png} \\
    %(b)
    \includegraphics[scale=0.42]{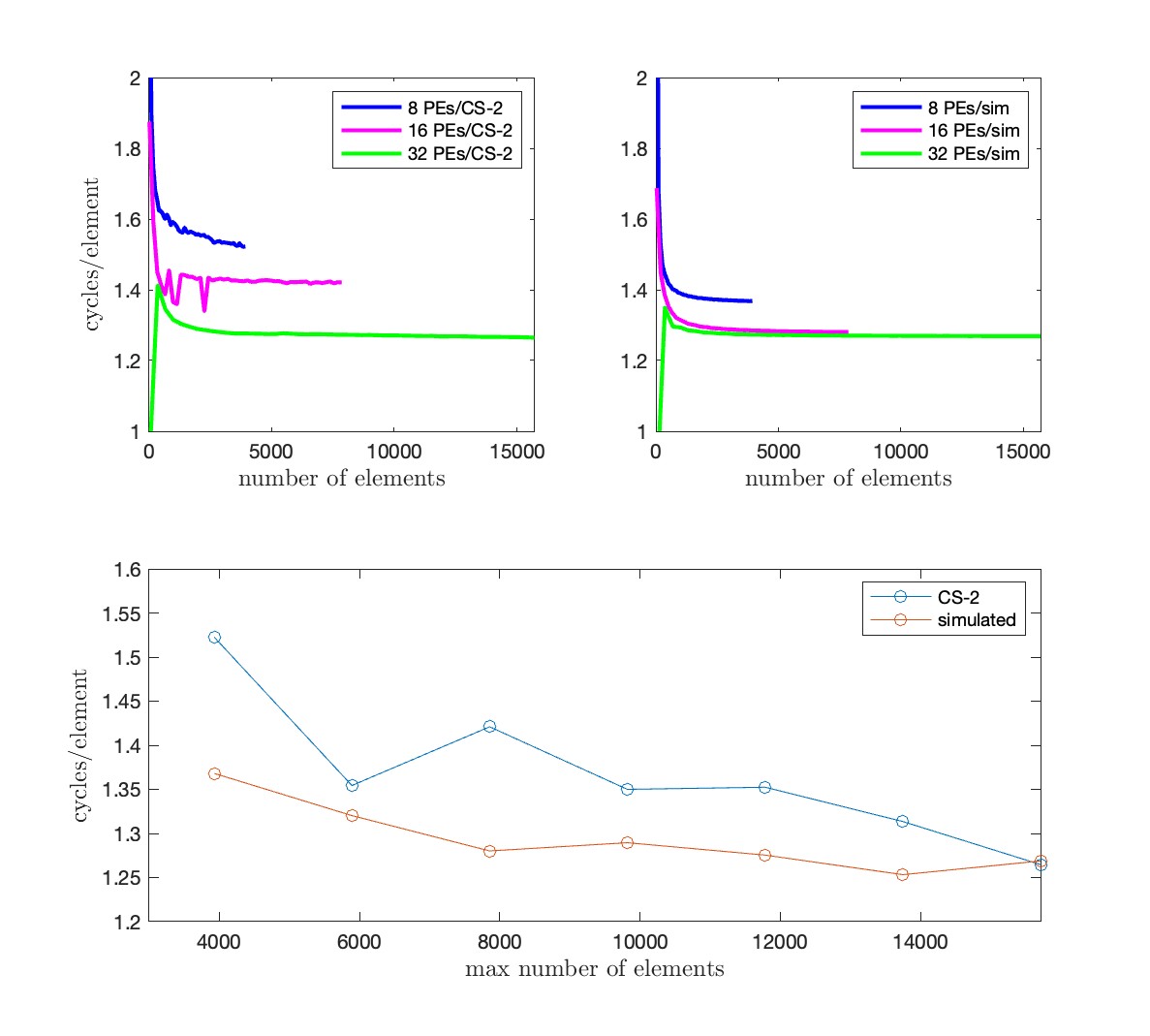}
\caption{
    Performance of Slide Eq. (\ref{EQN_s}) in cycles per array element versus total number of elements in individual benchmarks on the CS-2 (left top) versus simulated in the Cerebras SDK (right top). Data are distributed evenly over 8, 16, and 32 PEs, the number of elements per PE varying from 1 to 500. (Lower panel.) Results for maximal data arrays over $4k$ PEs ($k=2,3,\cdots, 8)$ shows performance to be essentially invariant to the number of PEs used by regular convergence to about 1.3 cycles per data element.
    }
    \label{figCS2}
\end{figure*}

On the CS-2, Eq. (\ref{EQN_s}) is implemented in the Cerebras Software Language (CSL) by the Cerebras Software Development Kit (SDK 1.0.0), enabling developers to write custom kernels for Cerebras systems.
CSL provides direct access to low-level hardware features while providing higher-level constructs such as loops and functions. The SDK provides a CS-2 simulator that allows software to be debugged and optimized without the physical wafer and provides cycle-accurate values for performance evaluation.

Fig.~\ref{figCS2} shows performance of our Slide operation Eq. (\ref{EQN_s}) in the fabric simulator and on a real CS-2 system. Realized is the desired linear scaling of slide-time with data-size in bytes and number of number of PEs involved.

Here, $\alpha=2/3$ given $a=2$ for a 64 bit datum in complex single precision, while typically $b=3$ including communication with Local Memory. Thus, (\ref{EQN_m}) is satisfied by an ample margin for typical FFT transform sizes.

\section{Conclusions}

The homogeneous mesh of PEs in WSEs offers a radically new architecture for low latency high-throughput signal processing. 
Deep searches for signals at small signal-to-noise ratios is frequently pursued by matched filtering evaluated in the Fourier domain by FFT. 

Seeking to exploit WSE, we propose compute-limited FFT by a new Slide operation Eq. \ref{EQN_s} 
facilitating arbitrary transform sizes. 
Performance Eq. (\ref{EQN_eta}) obtains by exploiting fast on-chip interconnect in Eqs. (\ref{EQN_eta}-\ref{EQN_m}).

Sliding data is required when transform sizes exceed the limitations of Local Memory per compute unit - the PE of a WSE. 
This approach circumvents the need for calls to Global Memory otherwise common in GPUs (Fig. \ref{fig0}), illustrative for the combined challenge of compute and memory overhead in seeking high throughput FFT-based signal processing on architectures different from CPUs.

Preliminary benchmarks demonstrate the desired linear scaling in Slide times with array size (Fig. \ref{fig:cs2arch}), 
essentially invariant to wave length - the number of PEs involved in the transform size. 
The proposed Slide operation hereby seems to provide a suitable starting point to ameliorate memory overhead in FFT on a homogeneous mesh architecture such as CS-2.

To exploit the full compute potential of CS-2 (240 TFLOPs), preserving the presented performance (Fig. \ref{figCS2}) to arbitrary transform sizes extended to batch mode further calls for limiting output back to the host device by post-callback functions, originally developed for implementation on GPUs \citep{van17}. With such additional steps in place to control memory overhead over the external bus, WSEs such as CS-2 hold realistic promise to accelerate the performance of high-throughput signal analysis by over an order of magnitude over conventional approaches. 

This outlook is of particular interest to searches for un-modeled signals by the FFT-based butterfly matched filtering algorithm \citep{van14}, nominally over 32 s segments of LIGO-Virgo-KAGRA data comprising $n=2^{17}$ samples \citep{van17}. This approach recently identified the physical trigger GW170817B - the central engine - of GRB170817A \citep{van19,van23a,van23b,abc23,van23c}. 

\mbox{}\\
{\bf Supplementary Data.} The CSL code of the Cerebras SDK generating the data of Fig. \ref{figCS2} is available on GitHub
https://github.com/leightonw-cerebras/sliding-window. 

\mbox{}\\
\begin{acknowledgments}
The authors gratefully acknowledge stimulating discussions with M.A. Abchouyeh.
%M. Hair and A.W. Lavely. 
%This is research received no external support.
\end{acknowledgments}


\appendix 

\section{Emulation of FFT}

\begin{figure}
    \centering
    \includegraphics[scale=0.22]{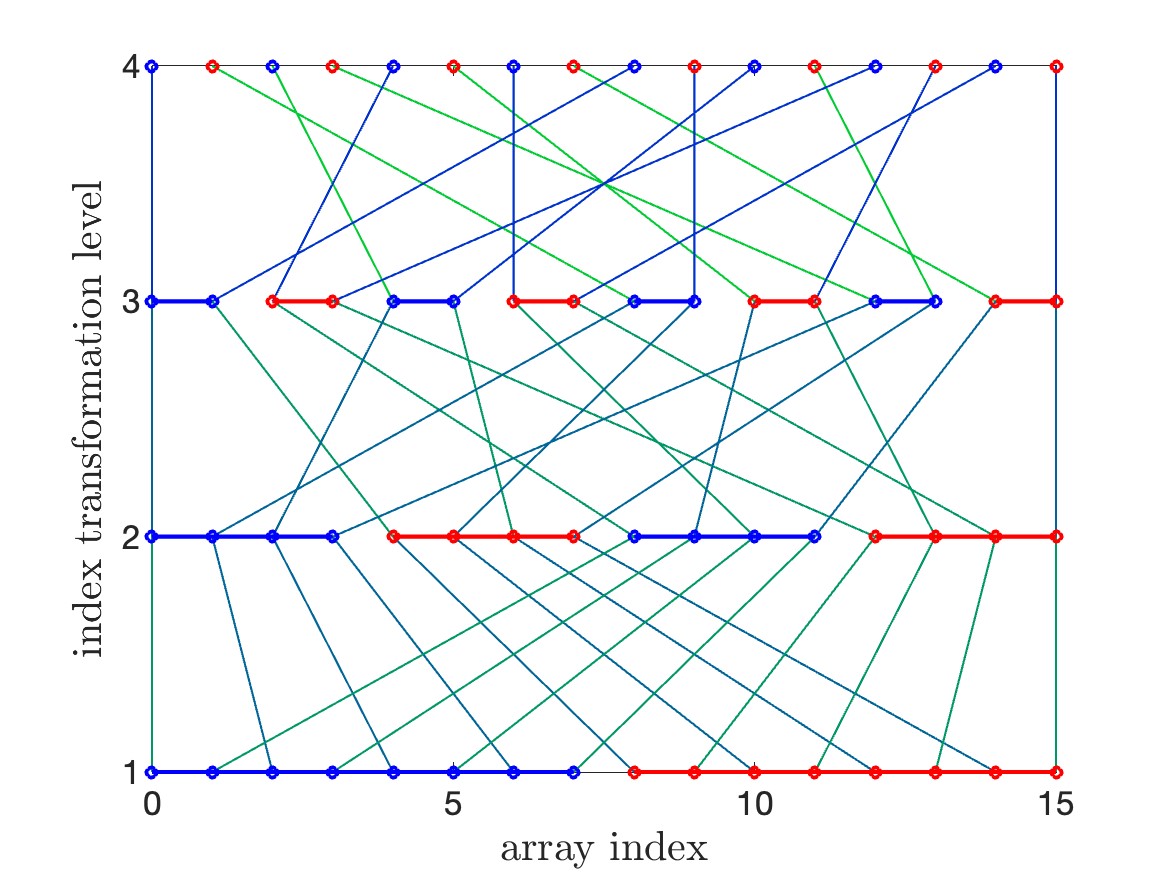}
    \caption{Array permutation of size $n=2^m$ ($m=4$ shown) over a partitioning of segments of size $n/2^{p-1}$ in even ($E$, blue) and odd ($O$, red) indexed elements, iterated over levels $p=1,2,\cdots m$ ($m=4$ shown).
    }
    \label{figA1}
\end{figure}

FFT is initialized by a permutation (Fig. \ref{figA1}) over levels $p=1,2,\cdots, m$ partitioned by segments of size $n/2^{p-1}$ into adjacent $I^\prime$ and $I^{\prime\prime}$ with even $(E)$ and, respectively, odd $(O)$ indices of $I$:
\begin{eqnarray}
{\small
\begin{array}{l}
I(1,:)=0:n-1\\
I(2,:)=[I^\prime(1,:) \,\,I^{\prime\prime}(1,:)]\\
\mbox{{\bf while}} \, m > 2 \\
  ~~ m=m/2\\
  ~~ N=n/m\\
  ~~ \mbox{{\bf for}}\, v=1:N\\
~~~~    M=(1+m(v-1)):mv\\
~~~~    I(k+1,M)=[I^\prime(k,M))\,\, I^{\prime\prime}(k,M)]\\
  ~~ \mbox{{\bf end}}\\
  ~~ k=k+1\\
\mbox{\bf end}\\
\mbox{clear}\, J \\
\mbox{\bf for}\, p=1:m \\
~~\mbox{\bf for}\, i=0:2^m-1 \\
~~~~    s=(I(p,:)==i) \\
~~~~    [s,a]=\mbox{\bf sort}(s,\mbox{\bf 'descend'}) \\
~~~~    J[(p,i+1)=a(1)-1 \\
~~\mbox{\bf end} \\
\mbox{\bf end}
\end{array}
\label{EQN_EA}
}
\end{eqnarray}

Subsequently, FFT is evaluated by crossings (\ref{EQN_LR}) over $m$ levels (\ref{EQN_LR}-\ref{EQN_Y}), illustrated in Eq. \ref{EQN_1a} for $m=3$ levels.
The potential of embarrassingly parallel computing follows from the following:
\begin{eqnarray}
{\small 
\begin{array}{ll}
{\bf Index\,level}\,m\\
~~I=J(m,:) \\
{\bf Data\, indexed \, for\,  level}\, m \\
~~ Y=J(1,I+1)^T \\
%{\bf FFT}\\
N=2\\
{\bf for}\, p=m:-1:1\\
~~  K=(0:N/2-1)^T\\
~~  Up={\bf exp}\left(-(2\pi/N)\sqrt{-1}K\right) \\
\begin{array}{|l}
~~{\bf for}\, v=1:n/N \\
~~~~    (M_1,M_2)\,{\bf index\, adjacent\,}E,O \\
~~~~      m_1=1+(v-1)N;m_2=m_1+N/2-1\\
~~~~      M_1=m_1:m_2;M_2=M_1+N/2 \\
        
~~~~    {\bf Fetch\,adjacent\,}E,O \\
~~~~     E=Y(M_1)\\
~~~~     O=Y(M_2)\\
~~~~     Op = Up.*O\\
~~~~    {\bf Rotate \,}E,O^\prime\\
%~~~~      [Op=Up.*O] \\
~~~~      L=E+Op\\
~~~~      R=E-Op \\
~~~~    {\bf Update\, adjacent \,}L,R\\
~~~~      Y(M_1)=L\\
~~~~      Y(M_2)=R\\
~~  {\bf end}
\end{array}\\
~~  N=2N \\
{\bf end}
\end{array}
\label{EQN_EB}
}
\end{eqnarray}
In (\ref{EQN_EB}), the loop $v=1:n/N$ (left bar) at levels $p$ is amenable to parallel processing on a homogeneous mesh of PEs.

%\section{Impl}
\end{document}